# Mid-Infrared Upconversion Imaging Under Different Illumination Conditions

**ZHENG GE,**[1, 2, #] **ZHAO-QI-ZHI HAN,**[1, 2, #] **YI-YANG LIU,**[3] **XIAO-HUA WANG,**[1, 2] **ZHI-YUAN ZHOU,**[1, 2, *] **FAN YANG,**[4] **YIN-HAI LI,**[1, 2] **YAN LI,**[1, 2] **LI CHEN,**[1, 2] **WU-ZHEN LI,**[1, 2] **SU-JIAN NIU,**[1, 2] **AND BAO-SEN SHI** [1, 2, *]

[1] *CAS Key Laboratory of Quantum Information, University of Science and Technology of China, Hefei, Anhui 230026, China*

[2] *CAS Center for Excellence in Quantum Information and Quantum Physics, University of Science and Technology of China, Hefei 230026, China*

[3] *School of Physical Science and Technology, Lanzhou University, Lanzhou 730000, China*

[4] *National Key Laboratory of Electromagnetic Space Security, Tianjin 300308, China*

[#] *These authors contribute equally in this work.*

[*] *zyzhouphy@ustc.edu.cn*

[*] *drshi@ustc.edu.cn*

Converting the medium infrared field to the visible band is an effective image detection method. We propose a comprehensive theory of image up-conversion under continuous optical pumping, and discuss the relationship between the experimental parameters and imaging field of view, resolution, quantum efficiency, and conversion bandwidth. Theoretical predictions of upconversion imaging results are given based on numerical simulations, which show good agreement with experimental results. In particular, coherent and incoherent light illumination are studied separately and the advantages and disadvantages of their imaging performance are compared and analysed. This work provides a study of the upconversion image detection performance of the system, which is of great value in guiding the design of the detection system and bringing it to practical applications.

The mid-infrared (MIR) spectral region contains the absorption/emission spectral positions of numerous molecules and structures and is often referred to as the chemical fingerprint spectral region for the analysis of the composition of substances [1]. In addition, this band is closely related to the thermal radiation of objects and contains atmospheric communication windows. As a result, MIR spectroscopy has fruitful applications in areas such as biomedicine [2], environmental monitoring [3], communications [4] and remote sensing [5], stimulating extensive research in recent years. However, despite the many advantages of MIR spectroscopy and imaging, the development of detectors in this band is still unsatisfactory, limiting many practical applications. Compared to their visible or near-infrared (NIR) counterparts, MIR detectors suffer from low detection sensitivity, high noise levels, slow response times and high cost. The detectors based on low bandgap materials such as mercury cadmium telluride (MCT) and Indium antimonide (InSb) have achieved high sensitivity and quantum efficiency, but rely on deep cooling to reduce dark noise, which places an additional burden on the application.

Non-linear frequency upconversion techniques, where the signal light is converted to visible/near-infrared light and then detected using high-performance detectors based on wide bandgap materials (e.g. silicon), offer an effective alternative [6]. This technique is not new and has been relatively well developed for signal detection in the communications band [7], with the advantages of room temperature operation and real time processes. In recent years, upconversion detection research has begun to expand into the mid-infrared band, with studies emerging for single-mode and image detection[8-10]. Waveguide-based upconversion detectors (UCD) can easily increase power density to achieve high conversion efficiencies[11,12], but the perturbation of the spatial structure of the optical field makes them suitable only for single-point detection situations. Bulk crystals are often used as nonlinear media in upconversion image detection, and schemes based on ultrafast lasers[13], cavity enhancement[14], or continuous light with high-power[15] pumping have been reported. Another advantage of UCD over conventional semiconductor-based detectors is the flexible tuning of performance parameters such as bandwidth, field of view, resolution and efficiency, which also requires a systematic imaging theory to guide the experimental design. In previous research on this topic, the usual treatment has been to associate the imaging theory of a linear system with a non-linear variation process[16-18]. This type of approach is more convenient in providing resolved results of the final image, but also involves considerable approximations, including neglecting the effect of propagation evolution within the crystal. A better solution would be to combine the analytical derivation with numerical simulations, which can give more accurate predictions of the imaging results. Furthermore, previous research has focused more on the case of coherent light illumination, as lasers have been the mainstay of modern optical research since their inception due to their excellent properties. However, thermal imaging applications are more oriented towards the spontaneous radiation of the target, and therefore the study of UCD in non-coherent illumination scenarios is of great interest in terms of passive detection. Our work provides a derivation of universally applicable upconversion imaging results for coherent/incoherent scenarios and provides experimental and theoretical control studies on parameters such as bandwidth, field of view, resolution and efficiency.

Upconversion imaging is based on sum frequency generation (SFG) in nonlinear crystals, where mid-infrared signal light with frequency $\omega_s$ is upconverted to $\omega_{SF}$ by pump light with frequency $\omega_p$. This process satisfies the law of energy conservation, i.e., $\hbar\omega_{SF} = \hbar\omega_s + \hbar\omega_p$, where $\hbar$ is Planck's constant. In order to achieve the highest frequency conversion efficiency, Quasi-phase-matching (QPM) techniques are often used to compensate for the phase mismatch and the polarization period is designed such that $\Delta\vec{k} = \vec{k}_s + \vec{k}_p + \vec{k}_\Lambda - \vec{k}_{SF}$, satisfying the momentum conservation condition. In most non-linear application scenarios, colinear matching is considered in order to achieve greater beam crossover and thus higher conversion efficiency. However, for the image conversion scene, the non-linear process will be a combination of the colinear and non-colinear cases, as the entire signal light field contains the individual components of incidence. By decomposing the phase mismatch wave vectors in the transverse and longitudinal dimensions, universal scalar form expressions can be written as:

$$\begin{cases} k_{SF}\cos\varphi = k_s\cos\theta - k_p - k_\Lambda \\ k_{SF}\sin\varphi = k_s\sin\theta \end{cases}, (1)$$

where $k_i = 2\pi n_i/\lambda_i$ $(i = p, s, SF)$ are the wave vectors of the pump light, signal light and sum-frequency light respectively, which depend on their respective wavelengths $\lambda_j$; the refractive index $n_j$ is given by the Sellmeier equation; and $\theta$ and $\varphi$ denote the angles of the signal light and sum-frequency light with respect to the direction of polarisation respectively. In a practical upconversion process, as the wavelength and direction of the pump light are usually constant, the wavelength and angle of incidence of the signal light, which are the independent variables in Eq. 1, determine the corresponding parameters of the upconverted light. At the same time, the presence of bivariate variables means that the conversion bandwidth and field of view of the system are correlated with each other and should be discussed together within the constraints of the phase matching condition. Due to the widespread use of lasers in optical experiments, past research has generally focused on the case of monochromatic light, constrained by strict phase matching conditions. As an example, the periodically polarised lithium niobate (PPLN) crystals, which are currently widely used to achieve infrared upconversion imaging, typically have conversion bandwidths in the order of a few nanometres and very narrow field-of-view angles[19,20]. To ameliorate these dilemmas, one solution is to add a tunable term to Eq. 1 by adjusting the crystal temperature or scanning the pumping wavelength[21,22], thereby extending the interval of understanding. However, this adds significantly to the complexity of the system and makes it difficult to meet the requirements of real-time imaging. Another solution is the use of chirped crystals, which have been used to achieve adiabatic nonlinear conversions with large phase-matching bandwidths, and image conversions with large fields of view have also been demonstrated[23-25]. In addition, methods to increase the conversion bandwidth by rotating the direction of incidence of the crystal or signal light have been reported, which can be classified as a hyperspectral imaging method. However, in broad-spectrum imaging scenarios, particularly where the infrared image to be converted originates from completely incoherent light emitted by the radiation of an object, the light field collected by the detection system contains a variety of frequency and spatial components. In this case, there is always a matching wavelength to be converted over a wide range of incidence angles, thus directly enabling image conversion over a large field of view and a large bandwidth. Figure 1 compares the upconversion of single-frequency and broad-spectrum signal light, using a single-period crystal and a chirped crystal, respectively. The pump light wavelength was fixed at 1064 nm and the crystal temperature was set at a uniform 30°C. Fig. 1(a) shows the variation in the angle of incidence corresponding to different signal light wavelengths while the crystal has a poled period of 21.4 μm. The use of chirped crystals results in a significant increase in the range of incident angles for single-frequency light, as shown in Fig. 1(b). In Fig. 1(c), it shows that there is a maximum value of the poled period in the up-conversion process by using PPLN at 30℃, which means there are two images with different wavelengths can be converted in a single poled period crystal.

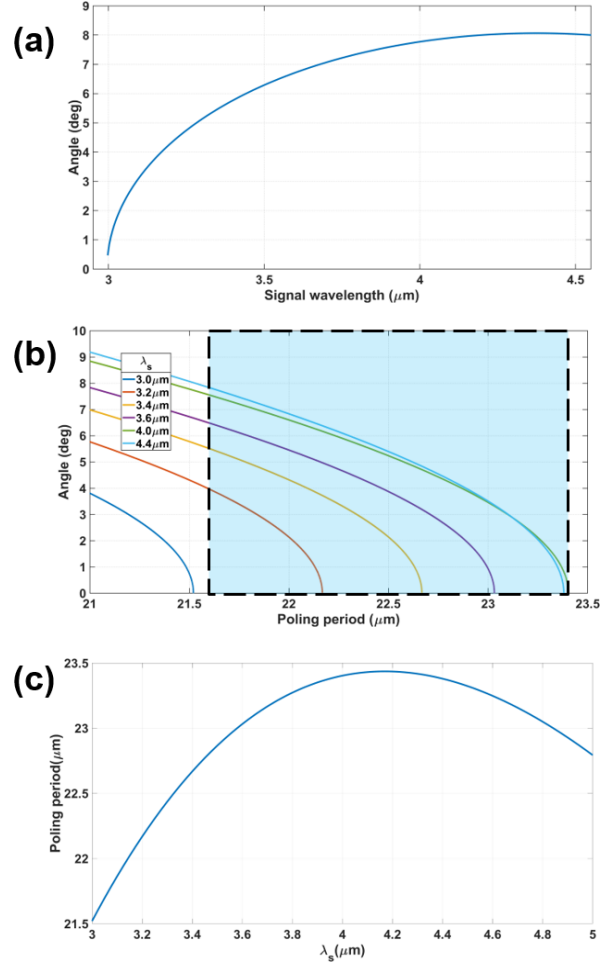

Fig. 1. (a) Theoretical calculation of input angle as a function of the MIR wavelengths (no-colinear case) MgO-PPLN crystal length is 30 mm, poled period is 21.5 μm (b) The optimized poled period for signal light with different wavelength and incident angle, the blue background area is the region that can meet phase matching condition (c) The optimized poled period for signal light with different wavelength and 0 degree angle, it can be seen that there is a maximum poled period which is calculated as 23.44 μm.

THEORETICAL MODEL AND SIMULATIONS

The general case of coherent light illumination is discussed first. There is a certain phase correlation between the points of the illuminated object, and the light maintains spatial and temporal coherence in the process of propagation. The propagation of coherent light illuminating objects in space can be described by the Collins formula, which is known as[26]

$$U(x,y) = \frac{\exp(ikl)}{i\lambda B} \iint U_0(x_0, y_0) \exp\left\{\frac{ik}{2B}\left[A(x_0^2 + y_0^2) + D(x^2 + y^2) - 2(xx_0 + yy_0)\right]\right\} dx_0 dy_0 , \quad (2)$$

Where, $A$, $B$, $C$, $D$ are the four elements in the transmission matrix; $U_0(x_0, y_0)$ and $U(x, y)$ represent the amplitude distribution of the illuminated object and the amplitude distribution after propagating $l$ on the $z$ axis.

For incoherent illumination, there is no spatial and temporal coherence instead. In the calculation process, the spatial coherence is eliminated by adding random phase fluctuations to the image and adding different propagation directions. The temporal coherence is eliminated by increasing the wavelength range of signal light and averaging multiple calculation results.

The wave equation of the electric vector propagating in the nonlinear crystal can be written as[27-29]:

$$\nabla^2 \vec{E}(\omega_n, z) + \mu_0 \omega_n^2 \varepsilon(\omega_n, z) \cdot \vec{E}(\omega_n, z) = -\mu_0 \omega_n^2 \overrightarrow{P_{NL}}(\omega_n, z), (3)$$

For second-order nonlinear processes, considering the slowly varying amplitude approximation, the scalar complex amplitude of the sum-frequency beam satisfies the differential equation, which gives

$$i2k_{SF}\frac{\partial E_{SF}}{\partial z} + \nabla_\perp^2 E_{SF} = -\frac{\omega_{SF}^2 d_{eff}}{\varepsilon_0 c^2} E_s E_p e^{i\Delta k z}, (4)$$

Where, $E_i (i = SF, s, p)$ represents the scalar complex amplitudes of sum-frequency beam, signal beam and pump beam, respectively; $\Delta k = k_{SF} - k_s - k_p + 2\pi/\Lambda$ is a phase mismatch during the sum frequency generation, $\Lambda$ is the poling period of the crystal; $\varepsilon_0$ and $c$ are the dielectric constant and the speed of light in the vacuum, respectively; $\nabla_\perp^2 = \frac{\partial^2}{\partial x^2} + \frac{\partial^2}{\partial y^2}$ represents the transverse Laplace operator.

The boundary conditions including the amplitude distribution of signal beam, pump beam, and sum-frequency beam at the front of the crystal can be easily obtained. By using the forward difference method to solve the sum frequency generation process in crystals, the coupled wave equation can be discretized as

$$E_{SF}(x, y, z+dz) = E_{SF}(x, y, z) + \frac{i}{2k_{SF}}\left[\nabla_\perp^2 E_{SF}(x, y, z) + \frac{\omega_{SF}^2 d_{eff}}{\varepsilon_0 c^2} E_s(x, y, z) E_p(x, y, z) e^{i\Delta k z}\right] dz, (5)$$

With this method, the amplitude distribution of three beams at any position in the crystal can be solved. Finally, the amplitude and intensity distribution of sum-frequency beam after imaging system can be calculated by using Collins formula again.

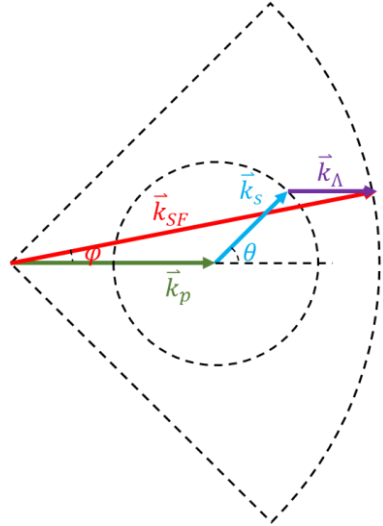

Fig. 2. Phase matching in sum-frequency generation for a certain wavelength of the signal beam. For coherent lighting objects, $\theta = 0$. For incoherent cases, $\theta$ can take any value. In chirped crystals, $\vec{k}_\Lambda$ is variable which can always find $\vec{k}_\Lambda$ that meets the phase matching conditions.

In the sum frequency generation, the propagation direction and wave vector of signal light are determined when the object is lighted by coherent beam. The propagation direction of the signal beam in the crystal is single. The lack of signal beam with large angle incidence leads that high-frequency information of its spatial structure cannot be converted, so its detailed imaging effect is not good. For incoherent light, its propagation direction within the crystal is random, and using chirped crystals can convert almost any direction of input beams. The wave vector direction that can be converted is only limited by the waist of the pump beam, which is generally smaller than the cross-sectional size of the crystal. Incoherent light illuminating objects can convert more directional wave vectors, so their high-frequency information can be well preserved, achieving clearer imaging effects (see Fig.2).

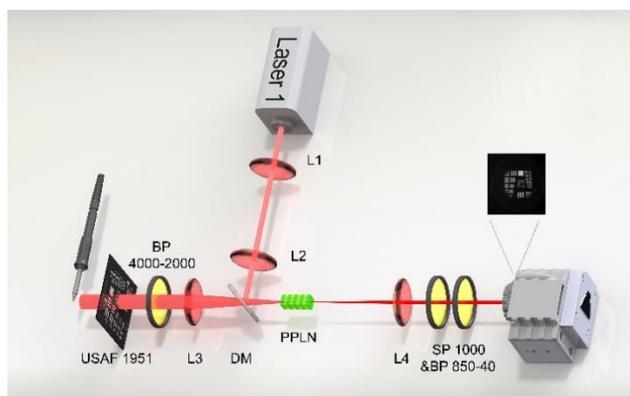

Fig. 3. Schematic diagram of the experimental setup. L terms: the lenses; DM: dichromatic mirror; BP: band-pass filter; PPLN: periodically poled lithium niobate crystal.

A simple diagram of our experimental setup is displayed in Fig. 3. We first generate the MIR source using a temperature adjustable soldering iron, and then use a collecting lens to transform the illumination light into a sub-parallel beam. The U.S. Air Force MIL-STD-150A standard of 1951 (USAF-1951) resolution target is then inserted as the illumination intensity object. The pump beam at the wavelength of 1064 nm comes from a Yb-doped fiber laser amplifier, and is scaled down to a 1/e beam waist of 1.5 mm through the lenses L1 ($f$ = 150 mm) and L2 ($f$ = 75 mm). The type-0 chirp PPLN crystal that we use in our experiment is 40 mm long, with an aperture of 2 × 3 $mm^2$, and its quasi-phase matching periodic poling period is from 21.6 to 23.4 μm, with an interval of 0.01 μm. The temperature of the PPKTP crystal is controlled using a homemade temperature controller with a stability of ±0.002 °C. Lens L3 ($f$ = 100 mm) focuses the MIR image into the PPLN crystal, which together with L4 ($f$ = 100 mm) forms the upconverted 4-f system, while the center of the PPLN crystal lies in the Fourier plane. The output image is cleaned up using a band-pass 800 nm filter with a full width at half maximum of 40 nm, and then recorded by an sCMOS camera (Dhyana95 V2, Tucsen).

In the above configuration, the central region of the USAF-1951 resolution card was selected as the intensity object, thus containing more resolution patterns in the field of view. Upconverted images of multiple sets of horizontal and vertical stripes were obtained and the output is shown in Figure 4(a). Figure 4(c) shows the numerical simulation results for a portion of the region where the upconversion system reaches its resolution limit. The advantage of this treatment is that it reduces the large computational effort that the oversized image imposes on the numerical simulation, and instead allows smaller step values to be assigned to the stepwise Fourier method, thereby increasing the accuracy of the simulation. Observe the right half of the experimental results, which agree with the theoretical and simulated results.

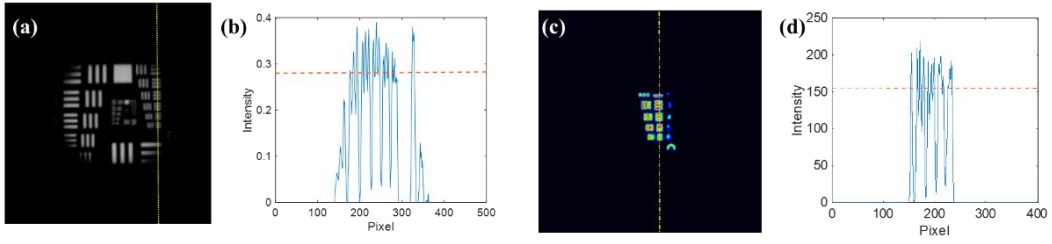

Fig. 4. Experimental incoherent up-conversion images of the USAF-1951 resolution card. (a) is the up-conversion image and (c) is the numerical simulation results. Line charts (b),(d) correspond to the vertical intensity distributions of images (a),(c), respectively.

For a specific analysis of the imaging resolution, we use the requirements of the Rayleigh criterion as an evaluation metric: $I_{min} \leq 2I_{max}/e$. Here $I_{max}$ is the average maximum intensity value of the line graph, while $I_{min}$ is the minimum grey value in the dark region between the stripes. Line graphs 4(b),(d) show the results of the vertical intensity distribution along the yellow dashed line from Fig. 4(a) to (c), where the orange horizontal dashed line is the resolution criterion given by the inequality in the above question. When the valley of the intensity in the streak region falls below the dashed line, this means that the set of patterns is resolvable. Experiments give limit resolution regions for groups 1-5 of the USAF-1951 resolution card, corresponding to a line resolution of 315 μm and an angular resolution of $3.15 \times 10^{-3}$ in our imaging system At the same time, simulations give a line resolution of 300 μm. It is worth noting that the theoretical angular resolution limit given by the system parameters is $R = 1.22\lambda_{MIR}/D_P$. Dp = 1.7mm is the beam diameter of the pump light inside the crystal; $\lambda_{MIR}$ is the wavelength of the signal light to be converted, here we take the center wavelength of 4.15μm within the conversion bandwidth Substitution of the relevant parameters gives R as $2.98\times10^{-3}$, which corresponds well to our experimental and simulations results. This also shows that our UCD system has been fully optimised and that the only factor limiting better resolution is the diaphragm limitation due to the aperture of the crystal.

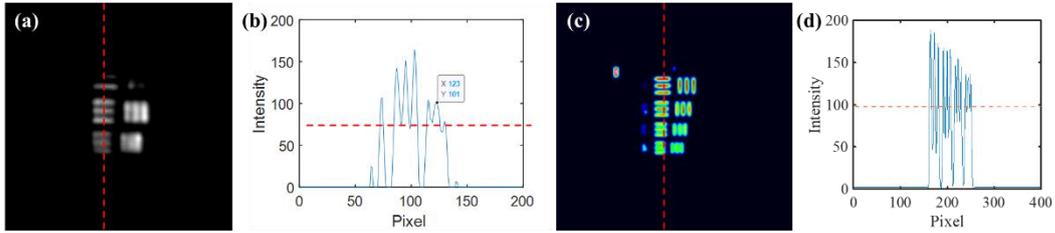

Fig. 5. Experimental coherent up-conversion images of the USAF-1951 resolution card. (a) is the up-conversion image and (c) is the numerical simulation results. Line charts (b),(d) correspond to the vertical intensity distributions of images (a),(c), respectively.

In coherent case, the experiments and simulations are shown in Fig.5. Both experiments and simulations give limited resolution regions for groups 1-4 of the USAF-1951 resolution card, corresponding to a line resolution of 353 μm. Compared to coherent cases, incoherent illumination has a better spatial resolution, which validates our theoretical analysis. In addition, due to the wide wavelength range of incoherent light generated by the thermal light source, its spectral range after SFG in the chirping crystal is also wider, which introduces dispersion and reduces spatial resolution.

For imaging experiments, a larger pump beam waist radius means a larger physical aperture, which improves imaging resolution by ensuring more conversion of the high frequency components. The thickness of the bulk crystal used in the experiments is 2 mm, which is currently the main factor limiting the optimisation of the system resolution. On the other hand, there is a reciprocal limitation between the resolution and the conversion efficiency of the system. This is because the use of long crystals to improve efficiency can lead to image distortion, while a large pump beam waist results in a significant reduction in power density. Therefore, the most effective way to improve the image resolution is increasing the waist of the pump beam as much as possible under the condition of acceptable conversion efficiency.

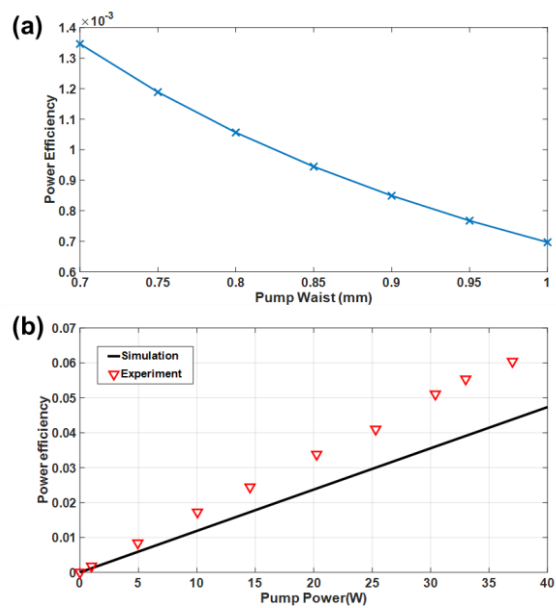

Fig. 6. Simulation and experimental results of upconversion efficiency. (a) Simulation results of the relationship between power efficiency and beam waist under the 1W pump. (b) Simulation and experimental results of the relationship between power efficiency and pump power at 0.75mm beam waist.

We also study the relationship between the waist of pump beam and the up-conversion efficiency of the image. The relationship between the upconversion efficiency and the beam waist and the pump light power is calculated, and the conversion efficiency under different pump optical power is measured experimentally. Through experiments, it was found that when the waist radius of the beam is 0.75mm, in the case of coherent light, the upconversion power efficiency is 0.184%, which is slightly higher than our calculated result 0.122%. It can also be verified through numerical calculations that the up-conversion efficiency is almost inversely proportional to the square of the waist radius of the pump beam. Both experiments and simulations give up-conversion power efficiency in incoherent cases 2-3 orders of magnitude lower than coherent illumination conditions. The reason is that incoherent illumination has random phase fluctuations and directions. It is necessary to design crystals, pump parameters and imaging systems for different application scenarios. In addition, the sensitivity of mid-infrared up-conversion imaging has also been studied. The intensity of the mid-infrared beam before the crystal is 0.813 pW. Under a 45W pump, the entire area count on the EMCCD camera is about $9.11 \times 10^6$ s$^{-1}$. The imaging results under the leak illumination are shown in Fig.7. In conclusion, increasing pump power not only improves conversion efficiency, but also reduces sampling time to achieve more sensitive detection. Q-switched lasers can also be used to achieve high efficiency, low noise and high sensitivity detection.

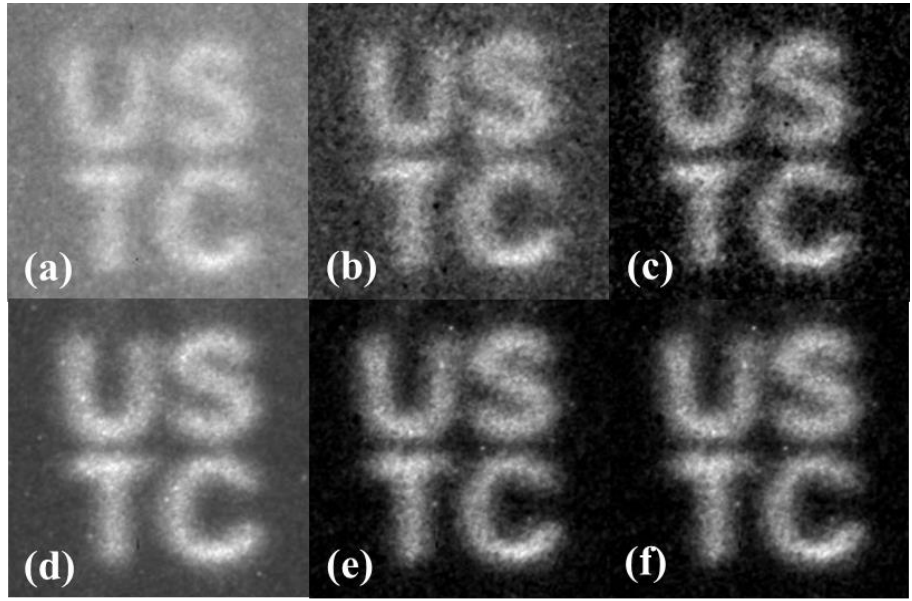

Fig. 7. (a) The noise is mainly composed of electrical noise and pump light noise, with a contrast of 1.74 while the image is not processed. (b) Subtracting the electrical noise result, the main noise is pump light noise, with a contrast of 3.00 (c) Subtracting electrical noise and pump light noise, the contrast is 27.88.(d)-(f) Imaging results of all noise, subtracting electric noise, subtracing electric noise and pump light noise under Q-switched laser pumping. The contrast are 2.85, 53.59 and 65.15, respectively.

At last, we change the crystal with different poled periods to investigate the field of the view, which is mainly limited by phase-matching condition in the experiment.We demonstrate that the crystals with a small poled period or large space inverted lattice vectors are beneficial for improving the field of view. In Fig. 8, we show the imaging results of two different chirp poled crystals under the same illumination conditions with the incoherent signal beam. When a chirp poled period crystal (with periods from 23.33 μm to 23.42 μm) is used to up-covert the image, it has a small field of view as shown in Fig.8 (b). This is because signal light with a large angle of incidence requires a smaller poled period to meet the phase matching condition witch is shown in Fig. 2. At this point, it can also be observed that when the crystal temperature changes, the field of view undergoes a significant change. The full angle of the field of view is 7.7 degree in this case.Then we chose the chirped crystals with periods from 21.6 μm to 23.4 μm, and the image under chirped this crystal has a larger field of view and higher intensity (see in Fig. 8(d)) than the case of the crystal with period from 23.33 μm to 23.42 μm. In this case, the full width of the imaging area is 29.5mm and the focal length of lens L1 is 100mm, so that full angle of the field of view is 16.8 degree. These results are consistent with the theoretical results shown in Fig. 1(b).

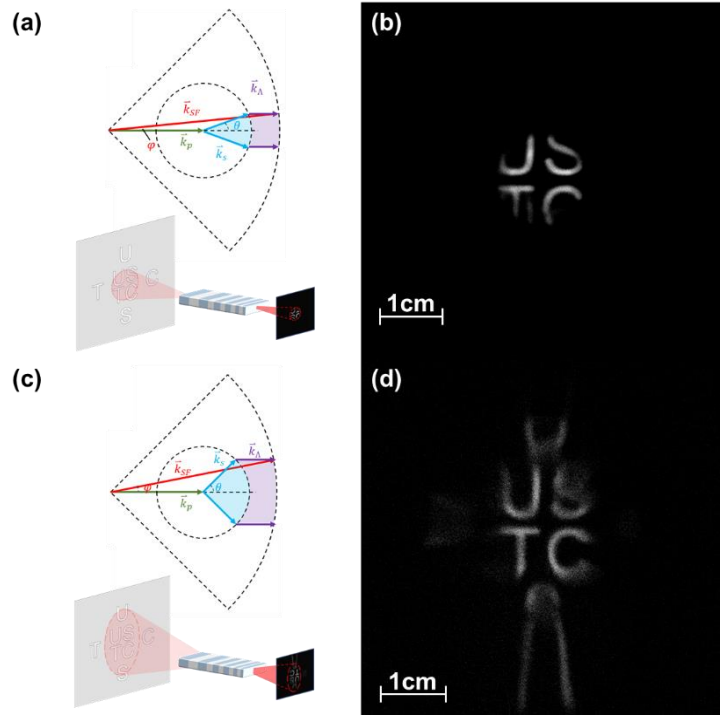

Fig. 8. (a) and (c) represent the imaging principles and schematic diagrams of the two different chirped poled crystals, respectively. The purple region represents the valuable range of spatial inverted lattice vectors in the chirped poled crystals, while the blue region represents the valuable range of the angle of signal beam input. The crystal with smaller poled periods has larger spatial inverted lattice vectors to achieve phase matching, thus corresponding to a larger angle of the field of view. (b) The image obtained by using the crystal with chirped poled periods from 23.33 μm to 23.42 μm. (d) The image obtained by using the crystal with chirped poled period of 21.6 μm to 23.4 μm.

In summary, we have investigated a MIR detector based on a frequency up-conversion process and its imaging properties. Mid-infrared upconversion detection in the presence of coherent and incoherent illumination has been achieved using single-period and chirped crystals, respectively. We theoretically analyse the imaging process of UCD, give theoretical predictions of the conversion results by numerical simulations and compare them with experimental results. The effects of experimental parameters on system efficiency, field of view and resolution are investigated, and the advantages of incoherent imaging over coherent processes are demonstrated. This work provides a systematic fundamental study of this frequency upconversion-based infrared imaging system, which will help to design and improve UCDs for a variety of application scenarios. By redesigning crystal parameters and imaging systems, this technology is expected to shine in astronomy, remote sensing, microscopy and other fields.

**Acknowledgments.** We would like to acknowledge the support from the National Key Research and Development Program of China (2022YFB3607700, 2022YFB3903102), National Natural Science Foundation of China (NSFC) (11934013, 92065101, 62005068), and Innovation Program for Quantum Science and Technology (2021ZD0301100), and the Space Debris Research Project of China (No. KJSP2020020202).

**References**

1. M. N. Abedin, M. G. Mlynczak, and T. F. Refaat, "Infrared detectors overview in the short-wave infrared to far-infrared for CLARREO mission," in M. Strojnik and G. Paez, eds. (2010), p. 78080V.
2. T. P. Wrobel and R. Bhargava, "Infrared Spectroscopic Imaging Advances as an Analytical Technology for Biomedical Sciences," Anal. Chem. **90**(3), 1444–1463 (2018).
3. S. M. Mintenig, I. Int-Veen, M. G. J. Löder, S. Primpke, and G. Gerdts, "Identification of microplastic in effluents of waste water treatment plants using focal plane array-based micro-Fourier-transform infrared imaging," Water Research **108**, 365–372 (2017).
4. F. Bellei, A. P. Cartwright, A. N. McCaughan, A. E. Dane, F. Najafi, Q. Zhao, and K. K. Berggren, "Free-space-coupled superconducting nanowire single-photon detectors for infrared optical communications," Opt. Express **24**(4), 3248 (2016).
5. B. M. Walsh, H. R. Lee, and N. P. Barnes, "Mid infrared lasers for remote sensing applications," Journal of Luminescence **169**, 400–405 (2016).
6. A. Barh, P. J. Rodrigo, L. Meng, C. Pedersen, and P. Tidemand-Lichtenberg, "Parametric upconversion imaging and its applications," Adv. Opt. Photon. **11**(4), 952 (2019).
7. S.-K. Liu, C. Yang, S.-L. Liu, Z.-Y. Zhou, Y. Li, Y.-H. Li, Z.-H. Xu, G.-C. Guo, and B.-S. Shi, "Up-Conversion Imaging Processing With Field-of-View and Edge Enhancement," Phys. Rev. Applied **11**(4), 044013 (2019).
8. M. Mancinelli, A. Trenti, S. Piccione, G. Fontana, J.S. Dam, P. Tidemand-Lichtenberg, C. Pedersen, and L. Pavesi, "Mid-infrared coincidence measurements on twin photons at room temperature," Nature Communications **8**, 15184 (2017).
9. Y. Wang, K. Huang, J. Fang, M. Yan, E. Wu, and H.P. Zeng, "Mid-infrared single-pixel imaging at the single-photon level, " Nature Communications **14**, 1073 (2023).
10. J. Tomko, S. Junaid, P. Tidemand-Lichtenberg, and W.T. Masselink, "Novel mid-infrared imaging system based on single-mode quantum cascade laser illumination and upconversion," in *2017 Conference on Lasers and Electro-Optics Europe & European Quantum Electronics Conference* (CLEO/Europe-EQEC, 2017), pp. 1-1.
11. T. W. Neely, L. Nugent-Glandorf, F. Adler, and S. A. Diddams, "Broadband mid-infrared frequency upconversion and spectroscopy with an aperiodically poled LiNbO_3 waveguide," Opt. Lett. **37**(20), 4332 (2012).
12. K.-D. Buchter, M.-C. Wiegand, H. Herrmann, and W. Sohler, "Nonlinear optical down- and up-conversion in PPLN waveguides for mid-infrared spectroscopy," in *CLEO/Europe - EQEC 2009 - European Conference on Lasers and Electro-Optics and the European Quantum Electronics Conference* (IEEE, 2009), pp. 1–1.
13. Q. Zhou, K. Huang, H. Pan, E. Wu, and H. Zeng, "Ultrasensitive mid-infrared up-conversion imaging at few-photon level," Appl. Phys. Lett. **102**(24), 241110 (2013).
14. J. S. Dam, P. Tidemand-Lichtenberg, and C. Pedersen, "Room-temperature mid-infrared single-photon spectral imaging," Nature Photon **6**(11), 788–793 (2012).
15. S. Junaid, S. Chaitanya Kumar, M. Mathez, M. Hermes, N. Stone, N. Shepherd, M. Ebrahim-Zadeh, P. Tidemand-Lichtenberg, and C. Pedersen, "Video-rate, mid-infrared hyperspectral upconversion imaging," Optica **6**(6), 702 (2019).

16. A.J. Torregrosa, E. Karamehmedovic, H. Maestre, M.L. Rico, and J. Capmany, "Up-Conversion Sensing of 2D Spatially-Modulated Infrared Information-Carrying Beams with Si-Based Cameras," Sensors **20**(12), 3610 (2020).
17. C. Pedersen, E. Karamehmedovic, J.S. Dam, and P. Tidemand-Lichtenberg, "Enhanced 2D-image upconversion using solid-state lasers," Optics Express **17**(23), 20885-20890 (2009).
18. J.S. Dam, C. Pedersen, and P. Tidemand-Lichtenberg, "Theory for upconversion of incoherent images," Optics Express **20**(2), 1475-1482 (2012).
19. Z.Y. Zhou, Y. Li, D.S. Ding, Y.K. Jiang, W. Zhang, S. Shi, B.S. Shi, and G.C. Guo, "Generation of light with controllable spatial patterns via the sum frequency in quasi-phase matching crystals," Scientific Reports **4**, 5650 (2014).
20. S. Junaid, J. Tomko, M.P. Semtsiv, J. Kischkat, W.T. Masselink, C. Pedersen, and P. Tidemand-Lichtenberg, "Mid-infrared upconversion based hyperspectral imaging," Optics Express **26**(3), 2203-2211 (2018).
21. H. Maestre, A. J. Torregrosa, C. R. Fernández-Pousa, and J. Capmany, "IR-to-visible image upconverter under nonlinear crystal thermal gradient operation," Optics Express **26**(2), 1133-1144 (2018).
22. K. Huang, J.N. Fang, M. Yan, E. Wu, and H.P. Zeng, "Wide-field mid-infrared single-photon upconversion imaging," Nature Communications **13**(1), 1077 (2022).
23. S.M.M Friis and L. Hogstedt, "Upconversion-based mid-infrared spectrometer using intra-cavity LiNbO3 crystals with chirped poling structure," Optics Letters **44**(17), 4231-4234 (2019).
24. A. Bostani, A. Tehranchi, and R. Kashyap, "Super-tunable, broadband up-conversion of a high-power CW laser in an engineered nonlinear crystal, " Scientific Reports **7**, 883(2017).
25. A. Barh, M. Tawfieq, B. Sumpf, C. Pedersen, and P. Tidemand-Lichtenberg, "Upconversion spectral response tailoring using fanout QPM structures," Optics Letters **44**(11), 2847-2850 (2019).
26. H. T. Yura and S. G. Hanson, " Optical beam wave propa-gation through complex optical systems," Journal of the Optical Society of America A **4**(10), 1931-1948 (1987).
27. R.W. BOYD, " Nonlinear Optics, " Orlando (USA): Academic Press, 2008.
28. Y.R. SHEN, "The Principles of Nonlinear Optics, " New York (USA): Wiley, 1984.
29. D.A. Kleinman, " Nonlinear Dielectric Polarization in Optical Media," Physical Review **126**(6), 1977-1979 (1962).